\documentclass[twocolumn,english,aps,pra,superscriptaddress,showpacs,tightenlines]{revtex4-1}
\usepackage{amsmath}
\usepackage{graphicx}
\usepackage{amssymb}
\usepackage{CJK}
\usepackage{color}

\begin{document}

\begin{CJK*}{GBK}{song}

\title{Controlling radiative dynamics of a giant $\Lambda$-type atom via interference induced by the vacuum of a waveguide}
\author{Ci-Ming \surname{Deng} }
\affiliation{Key Laboratory of Low-Dimension Quantum Structures and Quantum Control of Ministry of Education, Key Laboratory for Matter Microstructure and Function of Hunan Province, Hunan Research Center of the Basic Discipline for Quantum Effects and Quantum Technologies, Xiangjiang-Laboratory and Department of Physics, Hunan Normal University, Changsha 410081, China}
\affiliation{Institute of Interdisciplinary Studies, Hunan Normal University, Changsha, 410081, China}
\author{Ge \surname{Sun} }
\affiliation{Key Laboratory of Low-Dimension Quantum Structures and Quantum Control of Ministry of Education, Key Laboratory for Matter Microstructure and Function of Hunan Province, Hunan Research Center of the Basic Discipline for Quantum Effects and Quantum Technologies, Xiangjiang-Laboratory and Department of Physics, Hunan Normal University, Changsha 410081, China}
\affiliation{Institute of Interdisciplinary Studies, Hunan Normal University, Changsha, 410081, China}
\author{Jing \surname{Lu} }
\affiliation{Key Laboratory of Low-Dimension Quantum Structures and Quantum Control of Ministry of Education, Key Laboratory for Matter Microstructure and Function of Hunan Province, Hunan Research Center of the Basic Discipline for Quantum Effects and Quantum Technologies, Xiangjiang-Laboratory and Department of Physics, Hunan Normal University, Changsha 410081, China}
\affiliation{Institute of Interdisciplinary Studies, Hunan Normal University, Changsha, 410081, China}
\author{Lan \surname{Zhou}}
\thanks{Corresponding author}
\email{zhoulan@hunnu.edu.cn}
\affiliation{Key Laboratory of Low-Dimension Quantum Structures and Quantum Control of Ministry of Education, Key Laboratory for Matter Microstructure and Function of Hunan Province, Hunan Research Center of the Basic Discipline for Quantum Effects and Quantum Technologies, Xiangjiang-Laboratory and Department of Physics, Hunan Normal University, Changsha 410081, China}
\affiliation{Institute of Interdisciplinary Studies, Hunan Normal University, Changsha, 410081, China}

\begin{abstract}
We investigate the dynamics of a $\Lambda$-type giant atom (GA) whose both transition coupled to
the guided modes of a one-dimensional (1D) waveguide at two spatially separated points with
the GA initially excited and the electromagnetic (EM) modes of waveguide in vacuum. The spontaneous
emission properties of this GA is investigated by solving the delay-differential equation for
the amplitude of the 3GA in its excited state. Signatures of non-Markovian behavior is manifested
in a population trapping in the excited state of the GA in the regime where the distance $d$ of the
coupling points is smaller or comparable to the coherent length $L$ characterizing the width of the
emitted wave packet. And an exact Markovian dynamics is also found when $d\geq L$ via the inference
by adjusting the energy spacing and the inherent time delay besides the complex phases in the
atom-light coupling, matching the behavior of a small atom coupled to a waveguide.

\end{abstract}
\pacs{}
\maketitle

\end{CJK*}\narrowtext

\section{Introduction}
A central goal of modern quantum physics is efficient manipulation of the
interaction between quantum emitters (QEs) and a quantized electromagnetic
(EM) field. This is of particular importance for constructing large-scale
quantum networks~\cite{QN08,QN18} where the evolution and interaction of its
constituents can be controlled at the quantum level. The elementary process
of constituents is the interaction between a single photon and a single QE.
In free space, the QE-field coupling leads to an excited QE unavoidably
decaying towards its ground state by emitting a single photon to any of a
continuum characterized by different wave vectors. Boundaries and artificial
dimensional reduction tailor the mode density of the electromagnetic field,
spontaneous emission is either enhanced, inhibited or even completely
hindered\cite{CookPRA35,Alberpra46,Meystr56,Zoller66}. In this context,
cavity QED~\cite{CavityRPP69} setups have been extensively studied due to
the confinement of light into smaller volumes resulting in the strong
enhancement of single QE coupling to one preferred mode and/or suppression
of the coupling to all unwanted modes. In parallel to these cavity-based
settings, the coupling of QEs to waveguides, called waveguide-QED system,
has also raised a large interest as it enables the strong interaction of a
single or multiple atoms with a guided mode~\cite%
{RMP89(17),PR718(17),RMP95(23)}. QEs coupled with the 1D waveguides behave
as a gate for photons fly in the 1D waveguide, such as single-photon switches%
\cite{ShenPRL95,LanPRL101,ZhouOE30}, routers\cite%
{HoiPRL107,lanPRL111,LuPRA89,LuOE23,routPRA94,WeiPRA89,AhumPRA99,routPRR02},
and frequency converters\cite{LanPRA89,QFC108PRLS,QFC85PRAS,QFC96PRAL}.

With the great technological progress to the chip scale, the waveguide-QED
system has been generalized to studies of a QE nonlocally coupled to photons
in the waveguide\cite{ZLPRA85,LuOL49}. QEs with dimension comparable to or
larger than the wavelength of its emitted photon are called giant atom (GA)%
\cite%
{GAE130PRL,GAE120PRL,GAE105PRA,GAE109PRA,GAN103PRAY,lu6PRR,Wu109PRA,Yin106PRA,Wang111PRA}.
One of the paradigmatic models is a single GA coupled to a waveguide at
multiple spatially separated points. A hallmark of GA-waveguide system is
the nonexponential atomic decay, a non-Markovian character, due to the
self-interference effect induced by the separated coupling points, which has
been manifested in the form of dressed atom-photon bound states~\cite%
{GuoPRR2,wangPRA101,wangPRL126,LimPRA107,ShenPRA109,ShenPRA112,YangPRA111,sunPRA111}
and experimentally demonstrated with superconducting qubits coupled to
surface acoustic waves~\cite{NM15NP1123}. And the engineered complex phases
in the atom-light couplings~\cite{PRX13(23)023039}, which determines the
chirality of a GA~\cite{NM133PRL063603}, leads to nonreciprocal transport of
photons~\cite{SunPRA113} and a preferred direction of emitted radiation.
In this paper, we consider a $\Lambda $-type GA whose both transitions are
coupled to a waveguide at two spatially separated points. A three-level
emitter (3LE) is more suitable for long-lived information storage in comparison
with the two-level emitter (2LE), since it transfers the coherence of light
into the coherence between ground states. We concentrate on the spontaneous
emission of this three-level giant atom (3GA) in single excitation of this system.
A delay-differential equation for the amplitude of the 3GA in its excited
state is followed from the Schrodinger equation to describe the
self-interference with the waveguide by virtual photon exchanges. A photonic
bound state can be formed in a continuum produced by the destructive
interference effects by tuning the distance between the coupling points or
the energy spacing when the distance of the two coupling points is smaller
or comparable to the coherent length characterizing the width of the emitted
wave packet. The formation of the photonic bound state leads to a departure
from the exponential decay, a presentation of a non-Markovian character. We
also show that the 3GA can exhibit exact Markovian dynamics via the
interference between two propagating channels for photons even for
non-negligible time delays in addition to tuning the complex phases in the
atom-light couplings.
\begin{figure}[tbph]
\includegraphics[width=8 cm,clip]{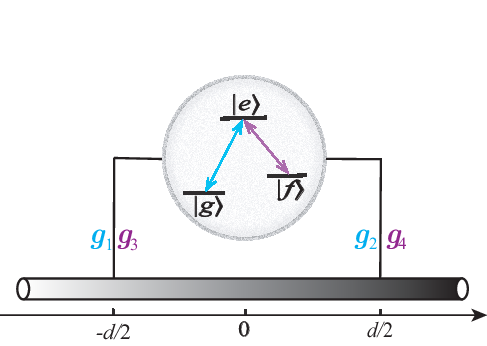}
\caption{(Color online) A sketch of the system: a three-level GA with a $%
\Lambda$-type configuration coupled to a linear waveguide. The upper level $%
\left\vert e\right\rangle $ is coupled to lower levels $\left\vert
g\right\rangle $ and $\left\vert f\right\rangle $ by the same vacuum modes
of a one-dimensional (1D) waveguide at position $x=-d/2$ and $d/2$. }
\label{figM}
\end{figure}

\section{\label{Sec:2}A GA Coupled to A 1D Waveguide}

We consider a 3GA in a so-called $\Lambda $ scheme with the upper
level $\left\vert e\right\rangle $, two lower levels $\left\vert
g\right\rangle $ and $\left\vert f\right\rangle $ separated by the
transition frequencies $\omega _{e}$ and $\omega _{ef}=\omega _{e}-\omega
_{f}$, respectively, as shown in the schematic Fig. 1. Note that $\omega
_{e}-\omega _{ef}=\omega _{f}$ is the frequency separation of the lower
levels with its ground energy being set to zero. The upper level $\left\vert
e\right\rangle $ is coupled to lower levels $\left\vert g\right\rangle $ and
$\left\vert f\right\rangle $ by the same vacuum modes of a one-dimensional
(1D) waveguide at position $x=-d/2$ and $d/2$. The waveguide supports a
continuum of electromagnetic modes, each with associated frequency $\omega $%
. Its 1D continuum is formed by the right-going and left-going modes, the
right-going mode are represented by the canonical creation and annihilation
operators $\hat{r}_{\omega }^{\dagger }$ and $\hat{r}_{\omega }$, and the
left-going mode are represented by $\hat{l}_{\omega }^{\dagger }$ and $\hat{l%
}_{\omega }$. Both modes obey the bosonic commutation relation $[\alpha
_{\omega },\alpha _{\omega ^{\prime }}^{\dagger }]=\delta (\omega -\omega
^{\prime })$ with $\alpha =\hat{r}_{\omega },\hat{l}_{\omega }$. The total
GA-waveguide Hamiltonian can be written as $\hat{H}=\hat{H}_{A}+\hat{H}_{W}+%
\hat{H}_{I}$, where
\begin{equation}
\hat{H}_{A}=\omega _{f}\hat{\sigma}_{ff}+\omega _{e}\hat{\sigma}_{ee}
\label{sec2-1}
\end{equation}%
is the Hamiltonian for the GA and atomic operator $\hat{\sigma}_{\beta \beta
^{\prime }}=\left\vert \beta \right\rangle \left\langle \beta ^{\prime
}\right\vert $ with $\beta ,\beta ^{\prime }\in \left\{ g,f,e\right\} $. The
free evolution of the quantized radiation field inside the waveguide is
described by
\begin{equation}
\hat{H}_{W}=\int_{0}^{\infty }d\omega \omega \left( \hat{r}_{\omega
}^{\dagger }\hat{r}_{\omega }+\hat{l}_{\omega }^{\dagger }\hat{l}_{\omega
}\right) .
\end{equation}%
By neglecting counter-rotating terms, the interaction of the GA with the
waveguide reads%
\begin{eqnarray}
\hat{H}_{I} &=&\frac{1}{\sqrt{\upsilon _{g}}}\int_{0}^{\infty }d\omega
\left( g_{1}e^{-\mathrm{i}\frac{\omega \tau }{2}}+g_{2}e^{\mathrm{i}\frac{%
\omega \tau }{2}}\right) \hat{r}_{\omega }^{\dagger }\hat{\sigma}_{ge}+H.c.
\label{sec2-2} \\
&&+\frac{1}{\sqrt{\upsilon _{g}}}\int_{0}^{\infty }d\omega \left( g_{3}e^{-%
\mathrm{i}\frac{\omega \tau }{2}}+g_{4}e^{\mathrm{i}\frac{\omega \tau }{2}%
}\right) \hat{r}_{\omega }^{\dagger }\hat{\sigma}_{fe}+H.c.  \notag \\
&&+\frac{1}{\sqrt{\upsilon _{g}}}\int_{0}^{\infty }d\omega \left( g_{1}e^{%
\mathrm{i}\frac{\omega \tau }{2}}+g_{2}e^{-\mathrm{i}\frac{\omega \tau }{2}%
}\right) \hat{l}_{\omega }^{\dagger }\hat{\sigma}_{ge}+H.c.  \notag \\
&&+\frac{1}{\sqrt{\upsilon _{g}}}\int_{0}^{\infty }d\omega \left( g_{3}e^{%
\mathrm{i}\frac{\omega \tau }{2}}+g_{4}e^{-\mathrm{i}\frac{\omega \tau }{2}%
}\right) \hat{l}_{\omega }^{\dagger }\hat{\sigma}_{fe}+H.c.  \notag
\end{eqnarray}%
where $g_{j}$ are GA-waveguide coupling strengths and $\tau =d/\upsilon _{g}$
is delay time that the light emitted from the GA into the waveguide travels
along the distance between two coupling points. As the total number of
excitations is preserved in this system, the subspace without excitations is
two-dimensional and is spanned by the bare states $\left\{ \left\vert
0g\right\rangle ,\left\vert 0f\right\rangle \right\} $ with $\left\vert
0\right\rangle $ being the vacuum state of the field. In the
single-excitation subspace, the state at time $t$ is a superposition of a
photon moving to the right or the left with the GA in any of the lower
state, and the GA is in the excited state
\begin{equation}
\begin{aligned}
\vert \Psi(t) \rangle
&= \sum_{\beta=g,f} \int_{0}^{\infty} d\omega \, \psi_{\beta l\omega}(t) \, l_{\omega}^{\dagger} \vert 0\beta \rangle
+ u(t) \vert 0e \rangle \\
&\quad + \sum_{\beta=g,f} \int_{0}^{\infty} d\omega \, \psi_{\beta r\omega}(t) \, r_{\omega}^{\dagger} \vert 0\beta \rangle
\end{aligned}
\label{sec2-3}
\end{equation}%
Here, $u\left(  t\right) $ is the excited amplitude, and $\psi _{\beta r\omega }\left(  t\right) $ ($\psi
_{\beta l\omega }\left(  t\right) $) are right- and left-moving field amplitudes with
frequency $\omega $ while the GA is in the state $\left\vert \beta
\right\rangle $. The Schr\"{o}dinger equation transforms $\left\vert \Psi
\left( t\right) \right\rangle $ into coupled equations of motion for the
amplitudes
\begin{subequations}
\label{sec2-4}%
\begin{align}
\dot{\psi}_{gl\omega}\left(  t\right)    & =-\mathrm{i}\omega\psi_{gl\omega
}\left(  t\right)  -\mathrm{i}\frac{u\left(  t\right)  }{\sqrt{\upsilon_{g}}%
}\left(  g_{1}e^{\mathrm{i}\frac{\omega\tau}{2}}+g_{2}e^{-\mathrm{i}%
\frac{\omega\tau}{2}}\right)  \\
\dot{\psi}_{gr\omega}\left(  t\right)    & =-\mathrm{i}\omega\psi_{gr\omega
}\left(  t\right)  -\mathrm{i}\frac{u\left(  t\right)  }{\sqrt{\upsilon_{g}}%
}\left(  g_{1}e^{-\mathrm{i}\frac{\omega\tau}{2}}+g_{2}e^{\mathrm{i}%
\frac{\omega\tau}{2}}\right)  \\
\dot{\psi}_{fl\omega}\left(  t\right)    & =-\mathrm{i}\left(  \omega
_{f}+\omega\right)  \psi_{fl\omega}\left(  t\right)  -\mathrm{i}\frac{u\left(
t\right)  }{\sqrt{\upsilon_{g}}}\left(  g_{3}e^{\mathrm{i}\frac{\omega\tau}%
{2}}+g_{4}e^{-\mathrm{i}\frac{\omega\tau}{2}}\right)  \nonumber\\
& \\
\dot{\psi}_{fr\omega}\left(  t\right)    & =-\mathrm{i}\left(  \omega
_{f}+\omega\right)  \psi_{fr\omega}\left(  t\right)  -\mathrm{i}\frac{u\left(
t\right)  }{\sqrt{\upsilon_{g}}}\left(  g_{3}e^{-\mathrm{i}\frac{\omega\tau
}{2}}+g_{4}e^{\mathrm{i}\frac{\omega\tau}{2}}\right)  \nonumber\\
& \\
\dot{u}\left(  t\right)    & =-\mathrm{i}\omega_{e}u\left(  t\right)
-\int_{0}^{\infty}d\omega\left(  g_{1}^{\ast}e^{\mathrm{i}\frac{\omega\tau}%
{2}}+g_{2}^{\ast}e^{-\mathrm{i}\frac{\omega\tau}{2}}\right)  \frac
{\mathrm{i}\psi_{gr\omega}\left(  t\right)  }{\sqrt{\upsilon_{g}}}\nonumber\\
& -\int_{0}^{\infty}d\omega\left(  g_{1}^{\ast}e^{-\mathrm{i}\frac{\omega\tau
}{2}}+g_{2}^{\ast}e^{\mathrm{i}\frac{\omega\tau}{2}}\right)  \frac
{\mathrm{i}\psi_{gl\omega}\left(  t\right)  }{\sqrt{\upsilon_{g}}}\nonumber\\
& -\int_{0}^{\infty}d\omega\left(  g_{3}^{\ast}e^{\mathrm{i}\frac{\omega\tau
}{2}}+g_{4}^{\ast}e^{-\mathrm{i}\frac{\omega\tau}{2}}\right)  \frac
{\mathrm{i}\psi_{fr\omega}\left(  t\right)  }{\sqrt{\upsilon_{g}}}\nonumber\\
& -\int_{0}^{\infty}d\omega\left(  g_{3}^{\ast}e^{-\mathrm{i}\frac{\omega\tau
}{2}}+g_{4}^{\ast}e^{\mathrm{i}\frac{\omega\tau}{2}}\right)  \frac
{\mathrm{i}\psi_{fl\omega}\left(  t\right)  }{\sqrt{\upsilon_{g}}}%
\end{align}
\end{subequations}
\begin{figure*}[tbp]
\includegraphics[width=0.8 \textwidth]{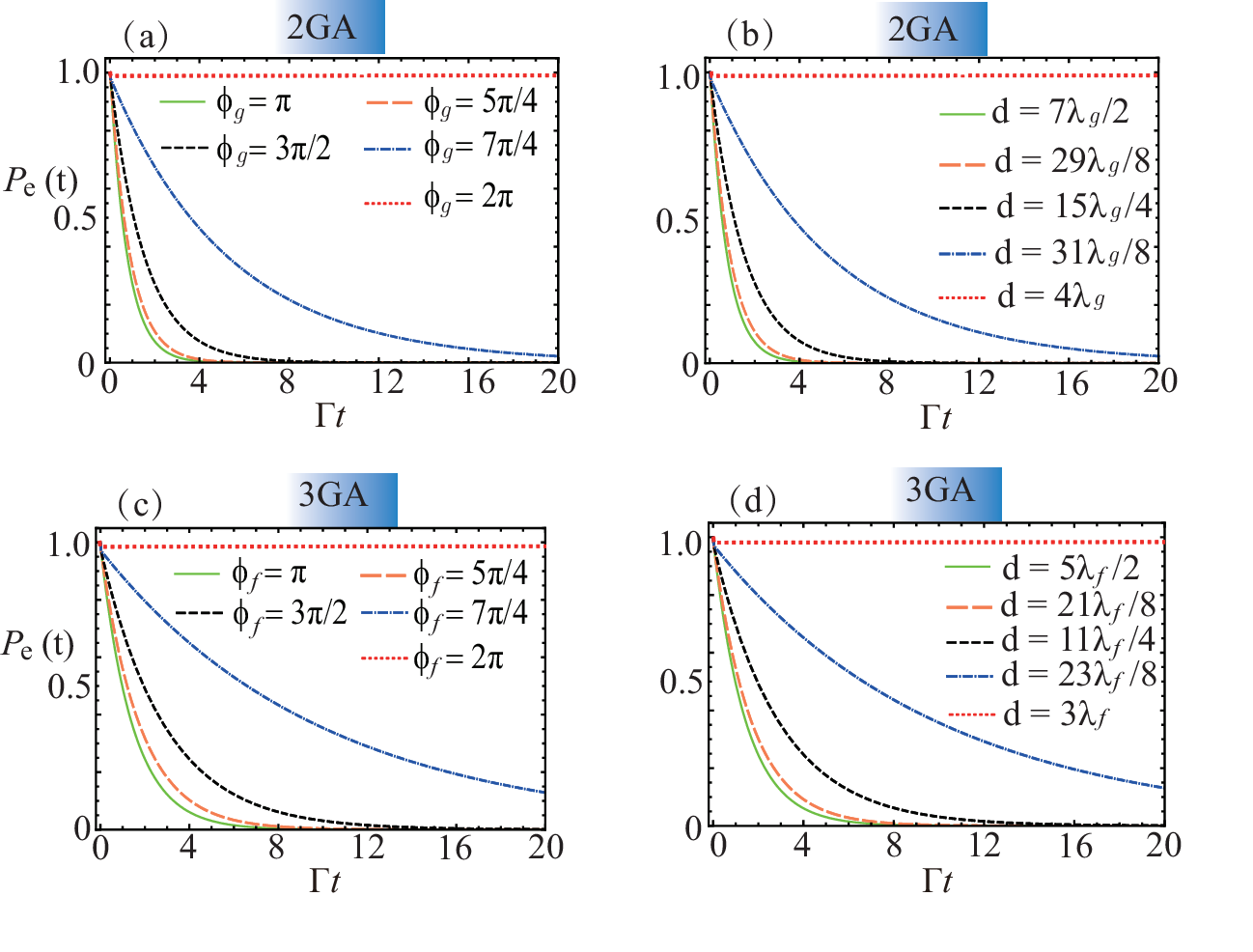}
\caption{(Color online) Time evolution of the population $P_e(t)$ for short
time delay $d\ll L$ with initial condition $u(0)=1$. The parameters are set
as follows: $\protect\omega_e=1400\Gamma$, $\Gamma_g=0.65\Gamma$, $%
|g_1|=|g_2|$ and $|g_3|=|g_4|$, (a) $\Gamma_f=0$, $d=7\protect\lambda_g/2$;
(b) $\Gamma_f=0$, $\protect\phi_g=\protect\pi$; (c) $\Gamma_f=0.35\Gamma$, $%
d=7\protect\lambda_g/2=5\protect\lambda_f/2$, $\protect\phi_g=2\protect\pi$;
(d) $\Gamma_f=0.35\Gamma$, $d=7\protect\lambda_g/2$, $\protect\phi_g=2%
\protect\pi$, $\protect\phi_f=\protect\pi$.}
\label{fig1}
\end{figure*}


\section{Atomic dynamics}

In this section we investigate the spontaneous emission of the GA initially
prepared in the excited state, i.e., the initial state is $\left\vert \Psi
\left( 0\right) \right\rangle =\left\vert 0e\right\rangle $. By formally
integrating the equations for $\psi _{\beta r\omega }\left( t\right)$ and $\psi _{\beta
l\omega }\left( t\right)$ with intial $\psi _{\beta r\omega }\left( 0\right) =\psi _{\beta
l\omega }\left( 0\right) =0$ and inserting them into the Eq.(\ref{sec2-4}e),
the photonic degree of freedom can be completely eliminated, yielding a
delay-differential equation
\begin{eqnarray}
\dot{U}\left( t\right) &=&-\left( \frac{\Gamma _{g}}{2}+\frac{\Gamma _{f}}{2}%
\right) U\left( t\right)  \label{sec3-01} \\
&&-\left( \frac{\gamma _{g}}{2}e^{\mathrm{i}\varphi _{g}}+\frac{\gamma _{f}}{%
2}e^{\mathrm{i}\varphi _{f}}\right) U\left( t-\tau \right)  \notag
\end{eqnarray}%
where we have introduced $u\left( t\right) =U\left( t\right) e^{-\mathrm{i}%
\omega _{e}t}$, $\Theta \left( t\right) $ is the Heaviside step function,
and
\begin{subequations}
\label{sec3-02}
\begin{eqnarray}
\Gamma _{g} &=&\frac{2\pi }{\upsilon _{g}}\left( \left\vert g_{1}\right\vert
^{2}+\left\vert g_{2}\right\vert ^{2}\right) , \gamma _{g} =\frac{4\pi }{%
\upsilon _{g}}\left\vert g_{1}\right\vert \left\vert g_{2}\right\vert \cos
\phi _{g}, \\
\Gamma _{f} &=&\frac{2\pi }{\upsilon _{g}}\left( \left\vert g_{3}\right\vert
^{2}+\left\vert g_{4}\right\vert ^{2}\right) , \gamma _{f} =\frac{4\pi }{%
\upsilon _{g}}\left\vert g_{3}\right\vert \left\vert g_{4}\right\vert \cos
\phi _{f}.
\end{eqnarray}%
The first term on the right hand side of Eq.(\ref{sec3-01}) corresponds to
the usual free space exponential decay with rate $\Gamma _{\alpha }$ to the
continua $\hat{l}_{\omega }^{\dagger }\left\vert 0\alpha \right\rangle $ and
$\hat{r}_{\omega }^{\dagger }\left\vert 0\alpha \right\rangle $, which are
referred to as $g$-channel and $f$-channel, respectively. The second term
represents the effect of the retarded radiation on the GA that was emitted
at time $\tau $ before it interacts again with the GA. Here, the emission
field of the GA acquires phases $\theta _{j}=\arg g_{j}$, $j\in \left\{
1,2,3,4\right\} $ at the coupling points generating the phase difference $%
\phi _{g}=\theta _{1}-\theta_{2}$ and $\phi _{f}=\theta _{3}-\theta _{4} $,
and accumulates propagation phases $\varphi _{g}=\omega _{e}\tau $ or $%
\varphi _{f}=\omega _{ef}\tau $ when light propagates from one coupling
point to its adjacent coupling point via $g$-channel or $f$-channel. The
time-dependent excited amplitude in Eq.(\ref{sec3-01}) can be calculated by
standard Laplace equations:
\end{subequations}
\begin{equation}
U\left( t\right) =\int \frac{e^{st}ds/\left( 2\pi \mathrm{i}\right) }{s+%
\frac{\Gamma }{2}+\frac{1}{2}\left( \gamma _{g}e^{\mathrm{i}\varphi
_{g}}+\gamma _{f}e^{\mathrm{i}\varphi _{f}}\right) e^{-s\tau }}
\label{sec3-03}
\end{equation}%
with the total decay rate $\Gamma =\Gamma _{g}+\Gamma _{f}$ and $s$ the
complex frequency parameter of the Laplace transformation. In the case of $%
\tau \rightarrow \infty $ or $g_{1}\left( g_{2}\right) =0$ and $g_{3}\left(
g_{4}\right) =0$, the system becomes a pointlike 3LE interacting with a
waveguide. When $g_{1}=g_{2}=0$ or $g_{3}=g_{4}=0$, the 3GA reduces to
a 2GA where the $g$-channel or $f$-channel disappears, and the 2GA further
reduces to a point-like 2LE when $g_{3}(g_{4})=0$ or $g_{1}(g_{2})=0$.
\begin{figure*}[tbp]
\includegraphics[width=0.8 \textwidth]{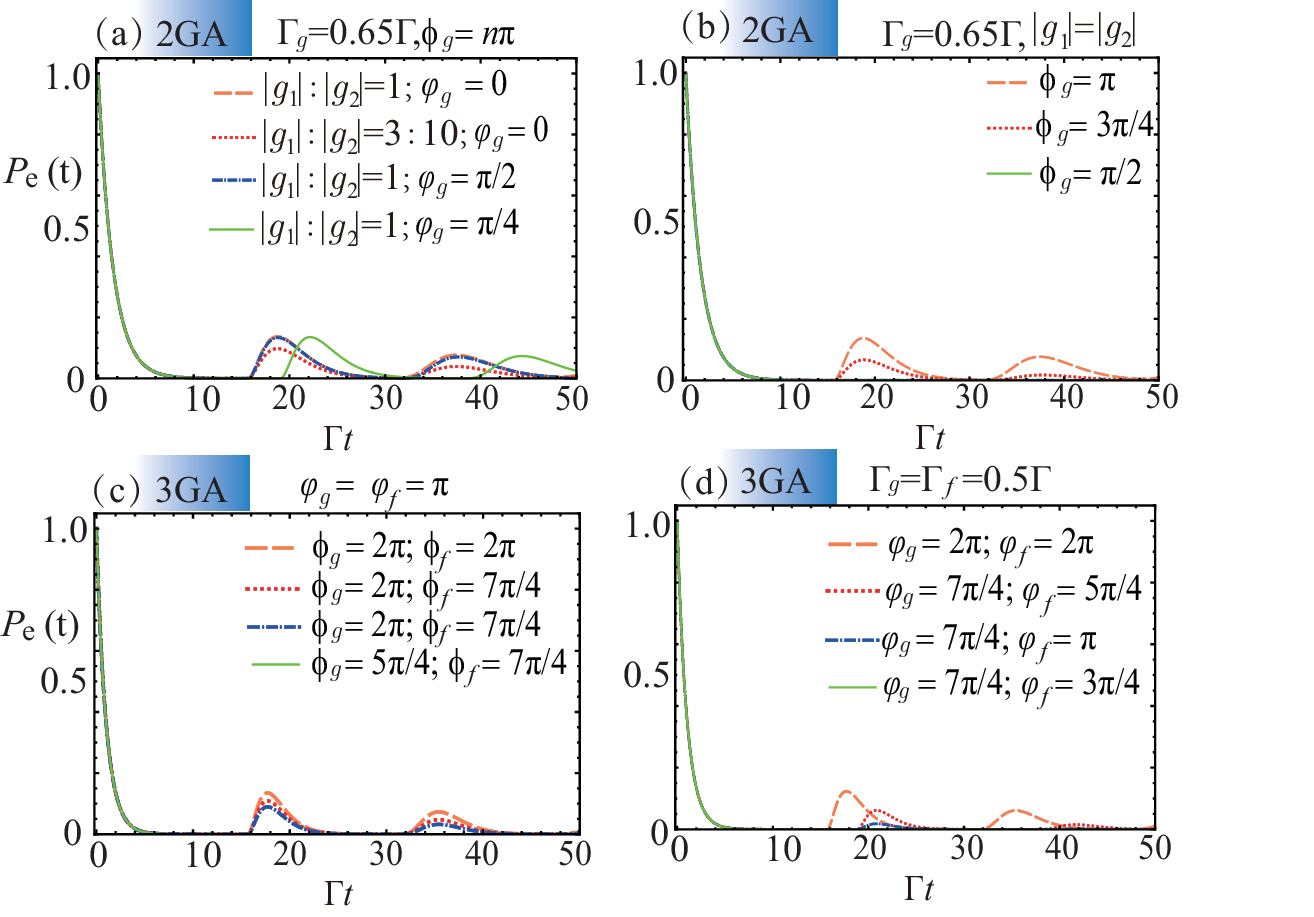}
\caption{(Color online) Time evolution of the population $P_{e}\left(
t\right)$ for the initially excited GA in long time delay. All curves are
obtained numerically. (a) $\protect\omega_e=1400\Gamma$ and $d=3500\protect%
\lambda_g$ for the orange dashed and red dotted lines; $\protect\omega%
_e=1400.2\Gamma, 1400\Gamma$ and $d=3500.5\protect\lambda_g, 4226.25\protect%
\lambda_g$ for the blue dot-dashed and the green solid line, respectively.
(b) $\protect\omega_e=1400\Gamma$, and $d=3500\protect\lambda_g$. (c) $%
\protect\omega_e=1400\Gamma$, $d=3500\protect\lambda_g=3500\protect\lambda_f$
are fixed, and $\Gamma_g=0.65\Gamma$, $\Gamma_f=0.35\Gamma$, $|g_1|=|g_2|$
and $|g_3|=|g_4|$ are setted for the orange dashed and red dotted lines; $%
\Gamma_g=\Gamma_f=0.5\Gamma$, $|g_1|:|g_2|=|g_3|:|g_4|=13:7$ and $|g_j|=g$
for the blue dot-dashed and the green solid lines, respectively. (d) $%
|g_1|:|g_2|=|g_3|:|g_4|=13:7$, $\protect\omega_e=1400\Gamma$ and $\protect%
\phi_g=\protect\phi_f=\protect\pi$ are fixed. The orange dashed, red dotted,
blue dot-dashed and green solid lines correspond to $d=3500\protect\lambda%
_g=2500\protect\lambda_f$, $d=(4200+7/8)\protect\lambda_g=(3000+5/8)\protect%
\lambda_f$, $d=(4200+7/8)\protect\lambda_g=(3000+1/2)\protect\lambda_f$ and $%
d=(4200+7/8)\protect\lambda_g=(3000+3/8)\protect\lambda_f$, respectively.}
\label{fig2}
\end{figure*}
Using geometric series expansion as $\gamma _{\beta }\leq \Gamma _{\beta }$
with $\beta \in \left\{ g,f\right\} $, the inverse Laplace transform yields
the time-dependent amplitudes
\begin{equation}
U\left( t\right) =\sum_{n=0}^{\infty }\frac{\left( -1\right) ^{n}}{2^{n}n!}%
\left( \gamma _{g}e^{\mathrm{i}\varphi _{g}}+\gamma _{f}e^{\mathrm{i}\varphi
_{f}}\right) ^{n}t_{n}^{n}e^{-\frac{\Gamma }{2}t_{n}}\Theta \left(
t_{n}\right)  \label{sec3-04}
\end{equation}%
where $t_{n}=t-n\tau $. Eq.(\ref{sec3-04}) has a \textquotedblleft
step\textquotedblright\ character if the time axis is divided into intervals
of length $\tau $. For the $n=0$ interval $t\in \left[ 0,\tau \right] $, the
sum consists only of one term, $\exp \left( -\Gamma t/2\right) $, coincides
with the behavior of a pointlike 3LE decaying to a 1D waveguide since the
emitted photon requires at least the time $\tau $ to travel between adjacent
coupling points. For $t\in \left[ \tau ,2\tau \right] $, the amplitude
consists of three terms since $n=1$ interval in Eq.(\ref{sec3-04}) is
included, which give rise to interference in the atomic population due to
the emitted radiation absorbed by the GA at the coupling point. And the
propagation of the emitted photon along $g$- and $f$-channel leads to the
dynamics of the GA strongly dependent on the propagating phases $\varphi
_{g} $ and $\varphi _{f}$. For longer times, more terms from the series
expansion (\ref{sec3-04}) contribute to the dynamics.

The role of the interference terms strongly depends on the distance $d$
between the coupling points, the wavenumbers $\lambda _{g}=2\pi \upsilon
_{g}/\omega _{e}$ and $\lambda _{f}=2\pi \upsilon _{g}/\omega _{ef}$, and
the coherence length $L=\upsilon _{g}/\Gamma $ describing the width of the
emitted wave packet. For GAs, $\lambda _{\beta }<d,L$ is usually satisfied.
In Fig.\ref{fig1}, we have plotted the probabilities $P_{e}\left( t\right)
=\left\vert u\left( t\right) \right\vert ^{2}$ in the excited state as a
function of the scaled time $\Gamma t$ in the short delay case under the
condition $\left\vert g_{1}\right\vert =\left\vert g_{2}\right\vert $ and $%
\left\vert g_{3}\right\vert =\left\vert g_{4}\right\vert $. The short delay
case corresponds to $d\ll L$, i.e., the time it takes for the GA to
completely decay to its two lower states is longer than the delay time. In
this context, the factor $e^{-s\tau }\approx 1$ in Eq.(\ref{sec3-03}), so a
Markovian approximation can be made. The exponential decay curves is usually
associated with the Markovian approximation. To show the difference of the
2GA and the 3GA, we set $g_{3}=g_{4}=0$ in Fig.\ref{fig1}(a,b). The behavior
of the black dashed line with $\phi _{g}=\left( n+1/2\right) \pi $ in Fig.%
\ref{fig1}(a) and $d=\left( n\pm 1/4\right) \lambda _{g}$ (i.e., $\varphi
_{g}=\left( 2n\pm 1/2\right) \pi $) in Fig.\ref{fig1}(b) is the same to the
exponential decay of the 2LE coupled to a 1D waveguide. The decay is either
enhanced or slowed down depending on the values of the parameters. The decay
vanishes at $\phi _{g}=2n\pi $ ($\left( 2n+1\right) \pi $) and $\varphi
_{g}=\left( 2n+1\right) \pi $ ($2n\pi $), and maximizes at $\phi _{g}=\left(
2n+1\right) \pi $ ($2n\pi $) and $\varphi _{g}=\left( 2n+1\right) \pi $ ($%
2n\pi $), see the red dotted lines and the green solid line in Fig.\ref%
{fig1}(a,b) respectively. The red dotted lines manifest a trapping of the
excitation in the 2GA. In Fig.\ref{fig1}(c,d), we have fixed $\phi _{g}=2\pi
$ and $\varphi _{g}=\left( 2n+1\right) \pi $ corresponding to complete
suppression of the spontaneous emission of the 3GA via the g-channel, see
the red dotted lines in Fig.\ref{fig1}(a). By allowing the transition $\left\vert e\right\rangle
\leftrightarrow \left\vert f\right\rangle $, the decay is generally
enhanced, and the behavior of the black dashed curves with $\phi _{f}=3\pi/2
$ in Fig.\ref{fig1}(c) and $d=\left( n\pm 1/4\right) \lambda _{f}$ (i.e., $%
\varphi _{f}=\left( 2n\pm 1/2\right) \pi $) in Fig.\ref{fig1}(d) is the same
to the exponential decay of the 2LA with transition $\left\vert
e\right\rangle $ and $\left\vert f\right\rangle $ coupled to a 1D waveguide.
The decay maximizes at $\phi _{f}=\left( 2n+1\right) \pi $ ($2n\pi $) and $%
\varphi _{f}=\left( 2n+1\right) \pi $ ($2n\pi $), see the green solid
curves. A trapping of the excitation in the excited state of the 3GA appears
at $\phi _{f}=2n\pi $ ($\left( 2n+1\right) \pi $) and $\varphi _{f}=\left(
2n+1\right) \pi $ ($2n\pi $), see the red dotted curves. In contrast to the
3GA with external driving field in Ref.\cite{YangPRA111,sunPRA111}, the
population trapping of the excitation is manifested only in the form of a
residual non-vanishing constant in the long enough time.

As the length of the distance of the coupling points grows such that $d\gg L$%
, the lifetime $\Gamma ^{-1}$ associated with the exponential decay factor
on each time interval is short compared with the duration $\tau $ of each
interval, the contributions from different intervals do not appreciably
overlap. In this long delay case, the probability reads
\begin{equation}
P_{e}\left( t\right) \approx \frac{1}{2^{2n}n!^{2}}\left\vert \gamma
_{g}+\gamma _{f}e^{\mathrm{i}\left( \varphi _{f}-\varphi _{g}\right)
}\right\vert ^{2n}t_{n}^{2n}e^{-\Gamma t_{n}}  \label{sec3-05}
\end{equation}%
in the $n$th time interval. It is possible for the GA to be partially
re-excited by the radiation which it has emitted before. Figure~\ref{fig2}
shows the probability $P_{e}\left( t\right) $ in the long delay case. The
exponential decay in the first interval ($0\leq t\leq \tau $) is followed by
a recurrence of probability beginning at $t=\tau $. The probability peaks at
$t=n\left( \tau +2/\Gamma \right) $ and reaches a peak value at
\begin{equation}
P_{e}\left( t\right) =\frac{1}{n!^{2}}\left\vert n\frac{\gamma _{g}+\gamma
_{f}e^{\mathrm{i}\left( \varphi _{f}-\varphi _{g}\right) }}{\Gamma
_{g}+\Gamma _{f}}\right\vert ^{2n}e^{-2n},  \label{sec3-06}
\end{equation}%
but decreases as $n$ increases. Consequently, the excitation is eventually
in the guided modes of the waveguide. The recurrence of probability always
begins at $t=n\tau $ since the photon completely leaves the GA before the
front of the wave packet reaches the other coupling point. For 2GA
originated from setting $g_{3}=g_{4}=0$ for the 3GA, it can be found that
the maximum peak value occurs when $\left\vert g_{1}\right\vert =\left\vert
g_{2}\right\vert $ for a given $\phi _{g}$ by comparing orange dashed line
and the red dotted line in Fig.\ref{fig2}(a). And the peak value further
increases until $\phi _{g}=n\pi $, see the orange dashed line and red dotted
line in Fig.\ref{fig2}(b). The change of frequency $\omega _{e}$ has no
effect on the probability $P_{e}\left( t\right) $ as the orange dashed line
and blue dot-dashed line are completely overlapping in Fig.\ref{fig2}(a).
When $\phi _{g}=\left( n+1/2\right) \pi $, the atomic decay is exponential
regardless of the time delay, see the green solid line in Fig.\ref{fig2}(b).
For the 3GA with degenerated two lower levels, $\varphi _{f}-\varphi _{g}=0$
in Eq.(\ref{sec3-05}), consequently, the energy spacing $\omega _{e}$ has no
effect, and the distance $d$ only determines the time of the recurrence.
Probability $P_{e}\left( t\right) $ has a peak value when $\left\vert
g_{1}\right\vert =\left\vert g_{2}\right\vert $ and $\left\vert
g_{3}\right\vert =\left\vert g_{4}\right\vert $ (see the difference between
the red dotted and the blue dot-dashed lines), and further reach its maximum
when $\phi _{g}=\phi _{f}=n\pi $ (see the difference between the orange
dashed line and the red dotted line). The 3GA can enters the Markovian
regime since we can make $\gamma _{g}+\gamma _{f}=0$ by appropriately
choosing $\phi _{g}$ and $\phi _{f}$ when $\left\vert g_{1}g_{2}\right\vert
=\left\vert g_{3}g_{4}\right\vert $, see the green solid line in Fig.\ref%
{fig2}(c). For the 3GA with non-degenerated two lower levels, $\varphi
_{f}-\varphi _{g}=\omega _{f}\tau $ gives the only difference from the
degenerated case, which means that he effect of the $|g_{j}|$ is similar to
the degenerated case but the energy spacing $\omega _{f}$ contributes to the
dynamics besides the distance $d$. Therefore, The behavior of the 3GA is
shown by fixing the ratio of $g_{1}$ $\left( g_{3}\right) $ to $g_{2}$ $%
\left( g_{4}\right) $ in Fig~\ref{fig2}(d). Here, the distance is changed
from $d=3500\lambda _{g}$ for the orange dashed line to $d=\left(
4200+7/8\right) \lambda _{g}$ for the red dotted line with the same $\omega
_{f}$, consequently, there is a shift of the revival time. The variation of
phase $\varphi _{f}$ for the red dotted line and the blue dot-dashed line is
adjusted by the frequency $\omega _{f}$. It can be found that the peak value
decreases (increases) as $\varphi _{f}-\varphi _{g}$ increases (decreases)
from $0$ ($\pi $) to $\pi $ ($0$) as $\phi _{g}=\phi _{f}$ ($\phi _{g}=\pi
+\phi _{f}$). It is worth pointing out that the Markovian dynamics can be
presented when one adjust $\omega _{f}$ to satisfy $\gamma _{g}+\gamma
_{f}e^{\mathrm{i}\omega _{f}\tau }=0$ for a 3GA, see the green solid line in
Fig~\ref{fig2}(d).

\begin{figure}[tbp]
\includegraphics[width=8 cm,clip]{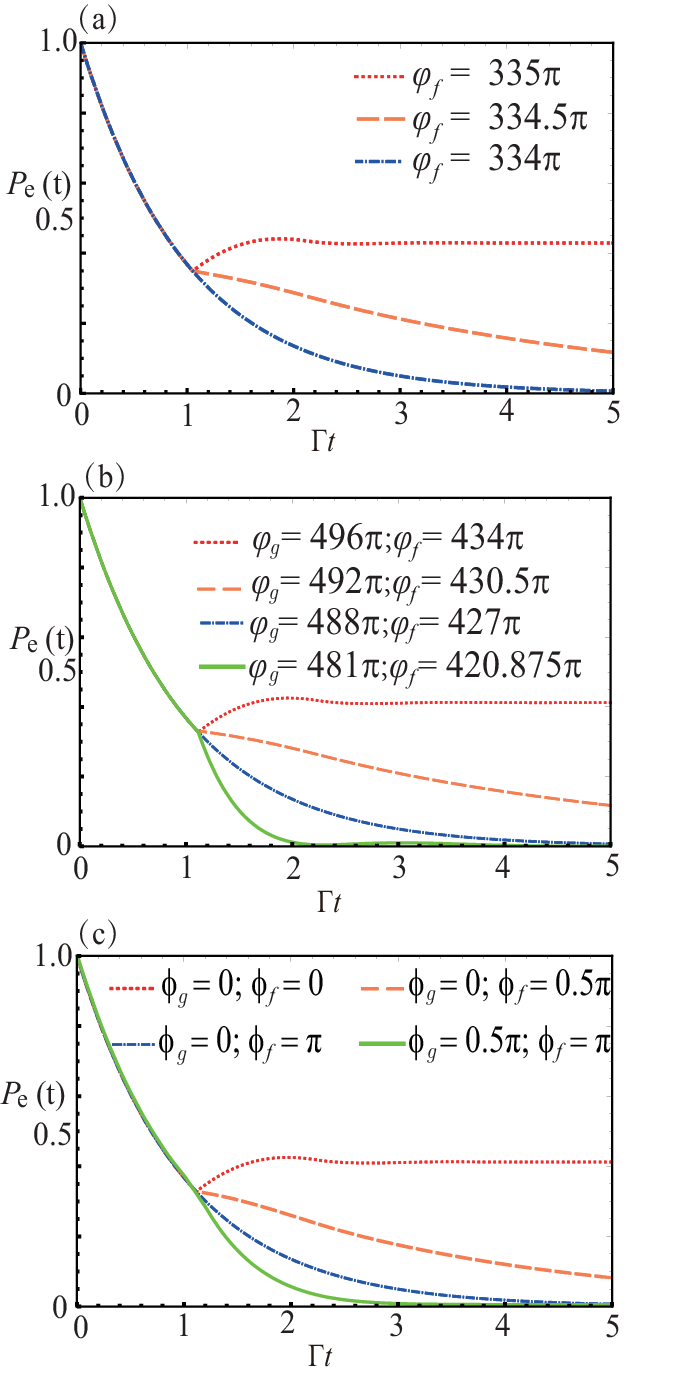}
\caption{(Color online) Time evolution of the excited-state population $%
P_{e}(t)$ for the initially excited 3GA with $d\sim L$. The parameters are
set as follows: $\protect\omega _{e}=1400\Gamma $, (a) $\protect\phi _{g}=%
\protect\phi _{f}=2\protect\pi $, $d=234.5\protect\lambda _{g}$. (b) $%
\protect\phi _{g}=\protect\phi _{f}=\protect\pi $, $\protect\lambda _{g}:%
\protect\lambda _{f}=7:8$. (c) $d=248.5\protect\lambda _{g}=177.5\protect%
\lambda _{f}$.}
\label{fig3}
\end{figure}

Having considered the short- and long-delay regimes, in the following, we
focus on an intermediate regime of $d$ comparable to $L$. In this
intermediate case, the previous intervals give more contribution to the next
interval. We let $|g_{j}|=cons.,j=1,\cdots ,4$ in order to give more concern
on the phases. In Fig.~\ref{fig3}, we show the dynamics of an initially
excited 3GA.
The excitation initially in the 3GA can be carried away by the guided modes.
One can adjust the propagating phase through frequency $\omega _{f}$ in
panel (a), the coupling distance $d$ in panel (b) and the coupling phase
differences $\phi _{\beta },\beta =g,f$ in panel (c) so that a 3GA presents
a dynamics exactly the same to a small QE coupled to a waveguide, see the
blue dot-dashed lines in Fig.~\ref{fig3}. A departure from the exponential decay is visible
after time $t>\tau $. The decay is enhanced (slowed down) comparing to that
of a small QE, see the green solid lines (orange dashed lines) in panels
(b,c). And it can be found that the parameters for the orange dashed lines
is due to the completely suppression of emission via the g-channel, in this
context, the dynamics at $t\geq \tau $ is exactly the same to the 2GA
coupled to a waveguide with transition $\left\vert e\right\rangle
\leftrightarrow \left\vert f\right\rangle $, subsequently, the GA is in
state $\left\vert f\right\rangle $ in the long time limit. It is worth
pointing out that 1) setting coupling phases $\phi _{\beta }=(n+1/2)\pi $
make a GA enter an exact Markovian regime irrespectively of any inherent
time delay, however, the propagating phases $\varphi _{\beta }=(n+1/2)\pi $
may result in an exponential decay over time only in the Markovian regime;
2) setting either $\phi _{f}-\phi _{g}=(2n+1)\pi $ when $\varphi _{g}=2n\pi
+\varphi _{f}$ or $\phi _{f}-\phi _{g}=2n\pi $ when $\varphi _{g}=(2n+1)\pi
+\varphi _{f}$ also makes a GA show a dynamics exactly the same to a small
QE due to the numerator null in Eq.(\ref{sec3-04}). A considerable trapping
of the probability within the 3GA for long times is clearly visible with a
steady value of population, see the red dotted lines. To underly the physics
of population trapping, let us find the poles of Eq.(\ref{sec3-04}). The
pure imaginary poles $s=-\mathrm{i}\omega $ require%
\begin{subequations}
\begin{eqnarray}
\omega &=&\frac{\gamma _{g}}{2}\mathrm{\sin }\left( \varphi _{g}+\omega \tau
\right) +\frac{\gamma _{f}}{2}\sin \left( \varphi _{f}+\omega \tau \right) ,
\\
-\frac{\Gamma }{2} &=&\frac{\gamma _{g}}{2}\cos \left( \varphi _{g}+\omega
\tau \right) +\frac{\gamma _{f}}{2}\cos \left( \varphi _{f}+\omega \tau
\right)
\end{eqnarray}
\end{subequations}%
to be satisfied. Solving these two equations, the real solution $\omega =0$
occurs for $\left\vert g_{1}\right\vert =\left\vert g_{2}\right\vert $ and $%
\left\vert g_{3}\right\vert =\left\vert g_{4}\right\vert $ with the four
following conditions for phases:
\begin{subequations}
\label{sec3-08}
\begin{eqnarray}
\phi _{g} &=&\phi _{f}=2n\pi ;\varphi _{g}=\varphi _{f}=\left( 2n+1\right)
\pi ; \\
\phi _{g} &=&\phi _{f}=\left( 2n+1\right) \pi ;\varphi _{g}=\varphi
_{f}=2n\pi ; \\
\phi _{g} &=&\varphi _{f}=2n\pi ;\phi _{f}=\varphi _{g}=\left( 2n+1\right)
\pi ; \\
\phi _{g} &=&\varphi _{f}=\left( 2n+1\right) \pi ;\phi _{f}=\varphi
_{g}=2n\pi .
\end{eqnarray}
\end{subequations}%
As the emission from the excited state to the guided modes are complete
suppressed, a dressed atom-photon bound state is formed.


\section{Field dynamics}

\begin{figure*}[tbp]
\includegraphics[width=0.8 \textwidth]{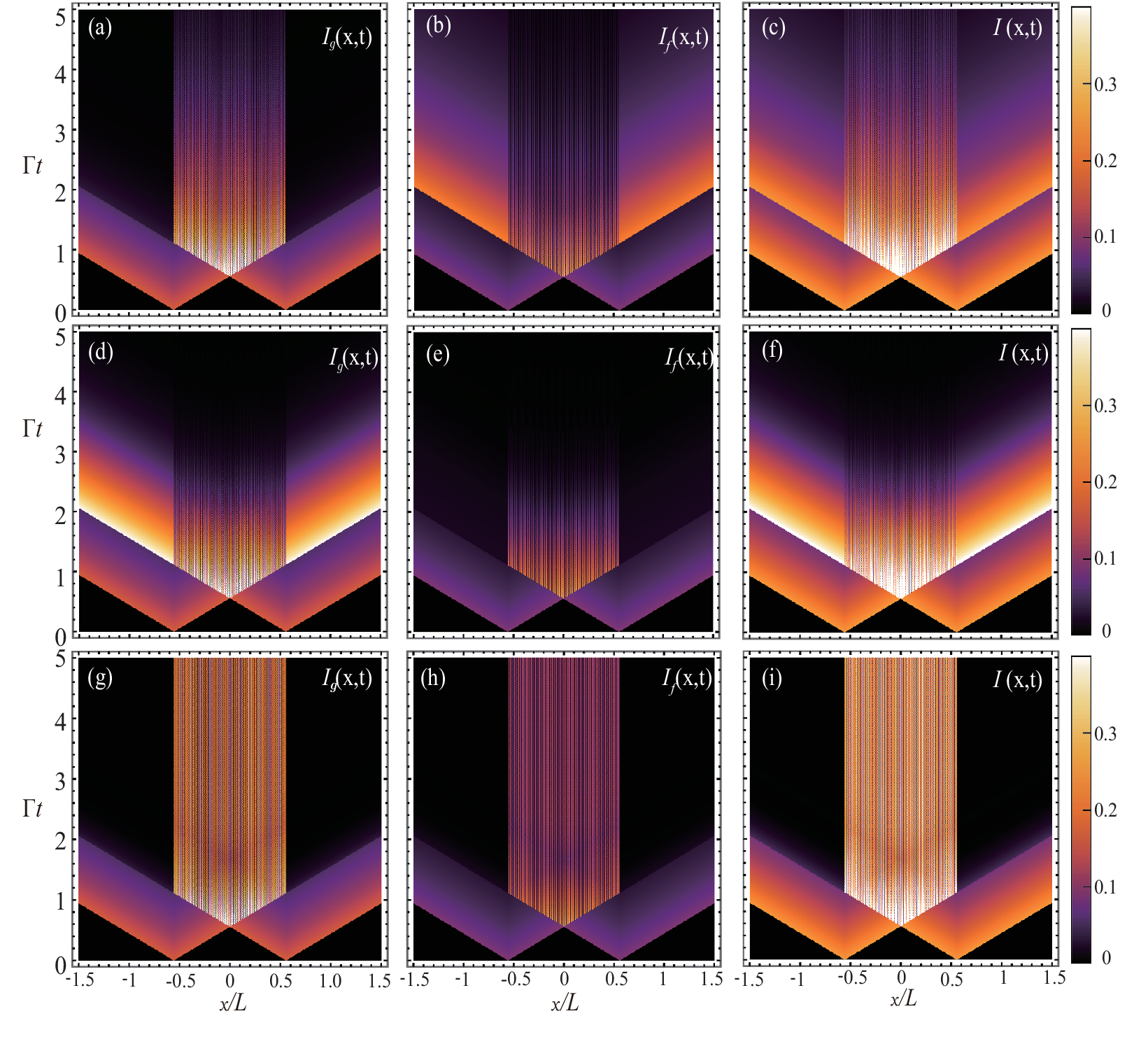}
\caption{The local photon densities $I_{g}(x,t)$, $I_{f}(x,t)$, and $I(x,t)$
with $u(0)=1$ for $d\sim L$. The parameters are set as follows: $\Gamma
_{g}=0.65\Gamma $, $\Gamma _{f}=0.35\Gamma $, $|g_{1}|=|g_{2}|$, $%
|g_{3}|=|g_{4}|$, $\protect\omega _{e}=1400\Gamma $, $d=248.5\protect\lambda %
_{g}=177.5\protect\lambda _{f}$. (a)-(c) $\protect\phi _{g}=\protect\pi ,%
\protect\phi _{f}=2\protect\pi $; (d)-(f) $\protect\phi _{g}=2\protect\pi ,%
\protect\phi _{f}=\protect\pi $; (g)-(i) $\protect\phi _{g}=\protect\phi %
_{f}=\protect\pi $.}
\label{fig4}
\end{figure*}
Spontaneous emission from an atom leads to a raise of the photon density in
the continuum of the waveguide. To get more physical insight as the radiation
field carries information about the atomic dynamics, we consider the local
photon density $I\left( x,t\right) =\left\langle \Psi \left( t\right)
\right\vert \hat{a}^{\dagger }\left( x\right) \hat{a}\left( x\right)
\left\vert \Psi \left( t\right) \right\rangle $ at position $x$ and time $t$%
, where the real-space field annihilation operator
\begin{equation}
\hat{a}\left( x\right) =\int_{0}^{\infty }\frac{d\omega }{\sqrt{2\pi
\upsilon _{g}}}\left( e^{-\mathrm{i}\frac{\omega x}{\upsilon _{g}}}\hat{r}%
_{\omega }+e^{\mathrm{i}\frac{\omega x}{\upsilon _{g}}}\hat{l}_{\omega
}\right)   \label{sec4-01}
\end{equation}%
is the sum of annihilation operators of the right- and left-going modes.
Obviously, $I\left( x,t\right)=I_{g}\left( x,t\right)+I_{f}\left( x,t\right)$ with $I_{\beta }\left( x,t\right)=\left\vert \Psi _{\beta
}\left( x,t\right)\right\vert ^{2}$ due to the presence of the $g$- and $f$-channel. And the
real-space field amplitudes are expressed as%
\begin{subequations}
\label{sec4-02}
\begin{align}
\Psi_{g}(x,t) &= \frac{1}{\sqrt{2\pi v_{g}}}
\int_{0}^{\infty} d\omega \left(
e^{-i\frac{\omega x}{v_{g}}}\psi_{gr\omega}\left(t\right)
+ e^{i\frac{\omega x}{v_{g}}}\psi_{gl\omega}\left(t\right)
\right), \\
\Psi_{f}(x,t) &= \frac{1}{\sqrt{2\pi v_{g}}}
\int_{0}^{\infty} d\omega \left(
e^{-i\frac{\omega x}{v_{g}}}\psi_{fr\omega}\left(t\right)
+ e^{i\frac{\omega x}{v_{g}}}\psi_{fl\omega}\left(t\right)
\right).
\end{align}
\end{subequations}
With the help of Eq.(\ref{sec2-4}) and (\ref{sec3-04}), the real-space
field amplitudes read
\begin{subequations}
\label{sec4-03}%
\begin{align}
\Psi_{g}\left(  x,t\right)    & =-\mathrm{i}\sqrt{\frac{\pi}{2}}\frac{g_{1}%
}{\upsilon_{g}}u\left(  t-\left\vert \tau_{x}^{+}\right\vert \right)
\Theta\left(  t-\left\vert \tau_{x}^{+}\right\vert \right)  \nonumber\\
& -\mathrm{i}\sqrt{\frac{\pi}{2}}\frac{g_{2}}{\upsilon_{g}}u\left(
t-\left\vert \tau_{x}^{-}\right\vert \right)  \Theta\left(  t-\left\vert
\tau_{x}^{-}\right\vert \right)  \\
\Psi_{f}\left(  x,t\right)    & =-\mathrm{i}\sqrt{\frac{\pi}{2}}\frac{g_{3}%
}{\upsilon_{g}}e^{-\mathrm{i}\omega_{f}\left\vert \tau_{x}^{+}\right\vert
}u\left(  t-\left\vert \tau_{x}^{+}\right\vert \right)  \Theta\left(
t-\left\vert \tau_{x}^{+}\right\vert \right)  \nonumber\\
& -\mathrm{i}\sqrt{\frac{\pi}{2}}\frac{g_{4}}{\upsilon_{g}}e^{-\mathrm{i}%
\omega_{f}\left\vert \tau_{x}^{-}\right\vert }u\left(  t-\left\vert \tau
_{x}^{-}\right\vert \right)  \Theta\left(  t-\left\vert \tau_{x}%
^{-}\right\vert \right)  \nonumber\\
& \null
\end{align}
\end{subequations}
where $\tau _{x}^{\pm }=x/\upsilon _{g}\pm \tau /2$. Figure~\ref{fig4}
numerically shows the dependence of the photon densities $I_{g}\left( x,t\right)$, $I_{f}\left( x,t\right)$
and $I\left( x,t\right)$ on time and coordinator for $d\sim L$. After a photon is emitted
from the coupling points at $\pm d/2$, it propagates towards left and right
respectively, which is also symbolized by the variables $t\pm x$ in Eq.(\ref%
{sec4-03}). Some waves propagate back and forth between the regime of the
coupling points. The emitted field interferes with itself and produce a
series of alternating bright and dark fringes for a period in panels (a-f).
During this period, the decay via the $g/f$-channel is completely
suppressed, see the dark area at $\left\vert x\right\vert \geq d/2$ in panel
(a/e), however, the spontaneous emission still proceeds via the $f/g$%
-channel, see the bright area at $\left\vert x\right\vert \geq d/2$ in panel
(b/d), the decay to $f/g$-channel leads the disappearance of the fringes in
the long time limit. Therefore, the total density in the waveguide are null
in the regime $\left\vert x\right\vert \leq d/2$ as photons propagate away
from the GA, see panels (c) and (f). However, the panels (g-i) in Fig.\ref%
{fig4} show an stationary interference pattern after an initial rapid change
with time, indicating that single photons are spatially localized in a
finite regime. This steady state of the field is a localized eigenmode with
energy eigenvalue residing directly in the continuum. The localized
eigenmode emerges due to the parameters in panels (g-i) satisfying the
condition in Eq.(\ref{sec3-08}). With the parameters in Eq.(\ref{sec3-08}),
the real-space field amplitudes in time $t>\Gamma $ can be further
simplified as
\begin{subequations}
\begin{align}
\Psi_{g}\left(  x,t\right)    & =\frac{\sqrt{2\pi}}{\upsilon_{g}}\frac
{g_{1}e^{-\mathrm{i}\omega_{e}t}\sin\left(  \omega_{e}\frac{2x+d}%
{2\upsilon_{g}}\right)  }{1+\frac{2\pi\tau}{\upsilon_{g}}\left(  \left\vert
g_{1}\right\vert ^{2}+\left\vert g_{3}\right\vert ^{2}\right)  }\\
\Psi_{f}\left(  x,t\right)    & =\frac{\sqrt{2\pi}}{\upsilon_{g}}\frac
{g_{3}e^{-\mathrm{i}\omega_{e}t}\sin\left[  \left(  \omega_{e}-\omega
_{f}\right)  \frac{2x+d}{2\upsilon_{g}}\right]  }{1+\frac{2\pi\tau}%
{\upsilon_{g}}\left(  \left\vert g_{1}\right\vert ^{2}+\left\vert
g_{3}\right\vert ^{2}\right)  }%
\end{align}
\end{subequations}
As the GA cannot fully decay to the two lower states, thus, less than one
photon exits the waveguide on average.

\section{Conclusion}

We study the dynamics of an upper level decaying to two lower states by
coupling to the guided modes of a linear waveguide in vacuum at two
spatially separated points in both Markovian and Non-Markovian regimes. The
delay-differential equation for the amplitude of the 3GA in its excited
state shows that two transitions induce two different propagating channels
for the emitted excitation in the waveguide, which introduces two
propagating independent phases $\varphi _{\beta },\beta =g,f$ on the
dynamics of the GA and results in the interference of the atomic population
in its excited state after reabsorbing its own emitted excitation at its
adjacent coupling point. In the Markovian regime, the $\Lambda$-type GA usually
decays exponentially, a departure from the exponential decay is presented by
a steady value for the dynamics of the GA in the long-time limit. In the
non-Markovian regime, the exponential decay is followed by a recurrence of
probability when the delay time is much larger than the lifetime of the
excited state. However, when $\tau \approx \Gamma ^{-1}$, its initial
exponential decay is followed by an enhanced, slow down of the decay and a
complete suppression of the emission of photons into the waveguide. The
analysis of the local density of the field shows that the quenching of
spontaneous emission is originated from the formation of the dressed
atom-photon bound state. We also found that the $\Lambda$-type GA can enter the
Markovian regime by the interference between two propagating channels in
addition to tuning the complex phases of the GA-photon couplings.

\begin{acknowledgments}
This work was supported by NSFC Grants No.12421005, No.12247105, XJ-Lab Key Project (23XJ02001),
and the Science $\And $ Technology Department of Hunan Provincial Program (2023ZJ1010).
\end{acknowledgments}


\begin{thebibliography}{99}
\bibitem{QN08} H. J. Kimble, The quantum internet, Nature (London) 453, 1023
(2008).

\bibitem{QN18} S. Wenhner, D. Elkouss, and R. Hanson, Quantum internet: A
vision for the road ahead, Science, 362, eaam9288 (2018).

\bibitem{CookPRA35} R. J. Cook, and P.W. Milonni, Quanum theory of an atom
near partial reflecting wall, Phys. Rev. A 35, 5081 (1987).

\bibitem{Alberpra46} G. Alber, Photon wave packets and spontaneous decay in
a cavity, Phys. Rev. A 46, R5338 (1992).

\bibitem{Meystr56} E. V. Goldstein and P. Meystre, Dipole-dipole interaction
in optical cavities, Phys. Rev. A 56, 5135 (1997).

\bibitem{Zoller66} U. Dorner and P. Zoller, Laser-driven atoms in
half-cavities, Phys. Rev. A 66, 023816 (2002).

\bibitem{CavityRPP69} H. Walther, B. T. Varcoe, B.-G. Englert, and T.
Becker, Cavity quantum electrodynamics, Rep. Prog. Phys. 69, 1325, (2006).

\bibitem{RMP89(17)} D. Roy, C. M. Wilson, and O. Firstenberg, Colloquium:
Strongly interacting photons in one-dimensional continuum, Rev. Mod. Phys.
89, 021001 (2017).

\bibitem{PR718(17)} X. Gu, A. F. Kockum, A. Miranowicz, Y.-X. Liu, and F.
Nori, Microwave photonics with superconducting quantum circuits, Phys. Rep.
718--719, 1 (2017).

\bibitem{RMP95(23)} A. S. Sheremet, M. I. Petrov, I. V. Iorsh, A. V.
Poshakinskiy, A. N. Poddubny, Waveguide quantum electrodynamics: Collective
radiance and photon-photon correlations, Rev. Mod. Phys. 95, 015002 (2023).

\bibitem{ShenPRL95} J. T. Shen and S. Fan, Coherent Single Photon Transport
in a One-DimensionalWaveguide Coupled with Superconducting Quantum Bits,
Phys. Rev. Lett. 95, 213001 (2005).

\bibitem{LanPRL101} L. Zhou, Z. R. Gong, Y.-X. Liu, C. P. Sun, and F. Nori,
Controllable Scattering of a Single Photon inside a One-Dimensional
Resonator Waveguide,\ Phys. Rev. Lett. 101, 100501 (2008).

\bibitem{ZhouOE30} J. L. Tan, X. W. Xu, J. Lu, and L. Zhou, Few-photon
optical diode in a chiral waveguide, Opt. Express 30, 28696 (2022).

\bibitem{HoiPRL107} I.-C. Hoi, C. M. Wilson, G. Johansson, T. Palomaki, B.
Peropadre, and P. Delsing, Demonstration of a Single-Photon Router in the
Microwave Regime, Phys. Rev. Lett. 107, 073601 (2011).

\bibitem{lanPRL111} L. Zhou, L. P. Yang, Y. Li, and C. P. Sun, Quantum
Routing of Single Photons with a Cyclic Three-Level System Phys. Rev. Lett.
111, 103604 (2013).

\bibitem{LuPRA89} J. Lu, L. Zhou, L. M. Kuang, and F. Nori, Single-photon
router: Coherent control of multichannel scattering for single photons with
quantum interferences, Phys. Rev. A 89, 013805 (2014).

\bibitem{LuOE23} J. Lu, Z. H. Wang, and L. Zhou, T-shaped single-photon
router, Opt. Exp. 23, 22955 (2015)

\bibitem{routPRA94} C. Gonzalez-Ballestero, E. Moreno, F. J. Garcia-Vidal,
and A. Gonzalez-Tudela, Nonreciprocal few-photon routing schemes based on
chiral waveguide-emitter couplings, Phys. Rev. A 94, 063817 (2016).

\bibitem{WeiPRA89} C. H. Yan, Y. Li, H. D. Yuan, and L. F. Wei, Targeted
photonic routers with chiral photon-atom interactions, Phys. Rev. A 97,
023821 (2018).

\bibitem{AhumPRA99} M. Ahumada, P. A. Orellana, F. Dom\'{\i}guez-Adame, and
A. V. Malyshev, Tunable single-photon quantum router, Phys. Rev. A 99,
033827 (2019).

\bibitem{routPRR02} B. Poudyal and I. M. Mirza, Collective photon routing
improvement in a dissipative quantum emitter chain strongly coupled to a
chiral waveguide QED ladder, Phys. Rev. Res. 2, 043048 (2020)

\bibitem{LanPRA89} Z. H. Wang, L. Zhou, Y. Li, and C. P. Sun, Controllable
single-photon frequency converter via a one-dimensional waveguide, Phys.
Rev. A 89, 053813 (2014).

\bibitem{QFC108PRLS} M. Bradford, K. C. Obi, and J.-T. Shen, Efficient
Single-Photon Frequency Conversion Using a Sagnac Interferometer, Phys. Rev.
Lett. 108, 103902 (2012);

\bibitem{QFC85PRAS} M. Bradford, J.-T. Shen, Single-photon frequency
conversion by exploiting quantum interference, Phys. Rev. A 85, 043814
(2012).

\bibitem{QFC96PRAL} W. Z. Jia, Y. W. Wang, and Y.-X. Liu, Efficient
single-photon frequency conversion in the microwave domain using
superconducting quantum circuits, Phys. Rev. A 96, 053832 (2017).

\bibitem{ZLPRA85} L. Zhou, Y. Chang, H. Dong, L.-M. Kuang, and C. P. Sun,
Inherent Mach-Zehnder interference with "which-way" detection for single
particle scattering in one dimension, Phys. Rev. A 85, 013806 (2012).

\bibitem{LuOL49} Z. L. Lu, J. Li, J. Lu, and L. Zhou, Controlling
atom-photon bound states in a coupled resonator array with a two-level
quantum emitter, Optics Letters, 49, 806 (2024).


\bibitem{GAE130PRL} A. C. Santos and R. Bachelard, Generation of Maximally
Entangled Long-Lived States with Giant Atoms in a Waveguide, Phys. Rev.
Lett. 130, 053601 (2023).

\bibitem{GAE120PRL} A. F. Kockum, G. Johansson, and F. Nori,
Decoherence-Free Interaction between Giant Atoms in Waveguide Quantum
Electrodynamics, Phys. Rev. Lett. 120, 140404 (2018).

\bibitem{GAE105PRA} A. Soro and A. F. Kockum, Chiral quantum optics with
giant atoms, Phys. Rev. A 105, 023712 (2022).

\bibitem{GAE109PRA} W. J. Gu, L. Chen, Z. Yi, S.J. Liu, and G.-X. Li, Tunable
photon-photon correlations in waveguide QED systems with giant atoms, Phys.
Rev. A 109 023720 (2024).

\bibitem{GAN103PRAY} L. Du, M. R. Cai, J. H. Wu, Z. H. Wang, and Y. Li,
Single photon nonreciprocal excitation transfer with non-Markovian retarded
effects, Phys. Rev. A 103, 053701 (2021).

\bibitem{lu6PRR} Q.-Y. Qiu and X.-Y. L\"{u}, Non-Markovian collective
emission of giant emitters in the Zeno regime, Phys. Rev. Res. 6, 033243
(2024).

\bibitem{Wu109PRA} S.-Y. Li, Z.-Q. Zhang, L. Du, Y. Li, and H.-Z. Wu, Single
photon scattering in giant-atom waveguide systems with chiral coupling,
Phys. Rev. A 109, 063703 (2024).

\bibitem{Yin106PRA} X.-L. Yin, W.-B. Luo, and J.-Q. Liao, Non-Markovian
disentanglement dynamics in double-giant-atom waveguide-QED systems, Phys.
Rev. A 106, 063703 (2022).
\bibitem{Wang111PRA} H. W. Yu, X. J. Zhang, Z. H. Wang, and J. Wang, Rabi oscillation and fractional population via the bound states in the continuum in a giant-atom waveguide QED setup, Phys. Rev. A 111, 053710 (2025).

\bibitem{GuoPRR2} L. Z. Guo, A. F. Kockum, F. Marquardt, and G. Johansson,
Oscillating bound states for a giant atom, Phys. Rev. Res. 2, 043014 (2020).

\bibitem{wangPRA101} W. Zhao and Z. H. Wang, Single-photon scattering and
bound states in an atom-waveguide system with two or multiple coupling
points, Phys. Rev. A 101, 053855 (2020).

\bibitem{wangPRL126} X. Wang, T. Liu, A. F. Kockum, H.-R. Li and F. Nori
Tunable chiral bound states with giant atoms Phys. Rev. Lett. 126 043602
(2021)

\bibitem{LimPRA107} K. H. Lim, W.-K. Mok, and L.-C. Kwek, Oscillating bound
states in non-Markovian photonic lattices, Phys. Rev. A 107, 023716 (2023).

\bibitem{ShenPRA109} Z. Y. Li and H. Z. Shen, Non-Markovian dynamics with a
giant atom coupled to a semi-infinite photonic waveguide, Phys. Rev. A 109,
023712 (2024).
\bibitem{ShenPRA112} S. J. Sun, Z. Y. Li, C. Cui, and H. Z. Shen, Non-Markovian dynamics of a driven three-level giant atom coupled to a semi-infinite waveguide via complex couplings, Phys. Rev. A 112,
063704 (2025).

\bibitem{YangPRA111} Y. Yang, G. Sun, J. Li, J. Lu, and L. Zhou, Coherent
Control of Spontaneous Emission for a giant driven $\Lambda$-type
three-level atom, Phys. Rev. A, 111, 013707 (2025).

\bibitem{sunPRA111} G. Sun, Y. Yang, J. Li, J. Lu, and L. Zhou,
Cavity-modified oscillating bound states with a $\Lambda $-type giant
emitter in a linear waveguide, Phys. Rev. A, 111, 033701 (2025)

\bibitem{NM15NP1123} G. Andersson, B. Suri, L. Guo, T. Aref, and P. Delsing,
Non-exponential decay of a giant artificial atom, Nat. Phys. 15, 1123 (2019).

\bibitem{PRX13(23)023039} C. Joshi, F. Yang, and M. Mirhosseini, Resonance
Fluorescence of a Chiral Artificial Atom, Phys. Rev. X, 13, 021039 (2023).


\bibitem{NM133PRL063603} F. Roccati, and D. Cilluffo, Controlling
Markovianity with Chiral Giant Atoms, Phys. Rev. Lett. 133, 063603 (2024).

\bibitem{SunPRA113} G. Sun, H. Z. Wu, J. Lu, and L. Zhou, Resonator-assisted
single-photon frequency conversion in a conventional waveguide with a giant
V-type atom, Phys. Rev. A, 113, 053719 (2026).
\end{thebibliography}
\end{document}